\documentclass[twocolumn,tighten,times]{aastex62}
\usepackage{color}
\usepackage{multirow}
\usepackage{chngcntr}
\usepackage{mathtools}
\usepackage[utf8x]{inputenc}
\usepackage{perpage}
\MakePerPage{footnote}
\usepackage{lipsum}
\usepackage[graphicx]{realboxes}
\makeatletter

\newcommand{\Rmnum}[1]{\expandafter\@slowromancap\romannumeral #1@}
\makeatother

\graphicspath{{./}{figures/}}

\received{\today}
\revised{tomorrow}
\accepted{the day after tomorrow}

\submitjournal{ApJ Letters}

\shorttitle{No or weak alignment of blue ETGs with filaments}
\shortauthors{Rong \& Wang}

\begin{document}

\title{Alignment of Blue/Green and Red Early-type Galaxies with Large-scale Filaments Reveals Distinct Evolutionary Pathways}

\correspondingauthor{Yu Rong}
\email{rongyua@ustc.edu.cn}

\author{Yu Rong}
%\altaffiliation{FONDECYT Postdoctoral Fellow}
\affiliation{Department of Astronomy, University of Science and Technology of China, Hefei, Anhui 230026, China}
\affiliation{School of Astronomy and Space Sciences, University of Science and Technology of China, Hefei 230026, Anhui, China}

\author{Peng Wang}
\affiliation{Shanghai Astronomical Observatory, Chinese Acedemy of Sciences, 80 Nandan Road, Shanghai 200030, China}

%%%%%%%%%%%%%%%%%%%%%%%%%%%%%%%%%%%%%%%%
\begin{abstract}
	
	We investigate the alignment of non-red early-type galaxies (ETGs) with blue or green colors within large-scale filaments and compare this alignment pattern with that of red ETGs. Our analysis reveals a significant alignment of the major axes of red ETGs with the orientations of their host cosmic filaments, consistent with prior research. In contrast, non-red ETGs show no significant alignment signal. This divergence in alignment behavior between non-red and red ETGs plausibly suggests distinct evolutionary pathways for non-red and red ETGs.

\end{abstract}


%%%%%%%%%%%%%%%%%%%%%%%%%%%%%%%%%%%%%%%%%

\section{Introduction}           %% first-level sections will be auto-capitalized
\label{sec:1}

Early-type galaxies (ETGs), encompassing primarily elliptical and lenticular galaxies, constitute a distinct cosmic population defined by spheroidal morphologies, older stellar populations, and typically low star formation activity. These galaxies are pivotal in exploring the connections between galaxy evolution and the larger cosmic environment. Studying the alignment of ETGs within large-scale structures like galaxy clusters, filaments, and voids provides insight into their formation processes, merger histories, and dynamical evolution \citep[e.g.,][]{Hung10, Primack24, OKane24, Merluzzi15}.

The spatial distribution and orientation of ETGs reveal alignment phenomena that inform our understanding of their assembly history and interactions with the cosmic web. Notably, ETG major axes often show a strong radial alignment within dynamically old, relaxed galaxy clusters and groups, likely due to angular momentum exchanges in repeated encounters within the dense cluster environment and tidal forces from central massive galaxies \citep{Schneider13, Rong15a, Rong19, Plionis08, Lee15, Coutts96, Pereira05, Wang17, 2018MNRAS.473.1562W}. Conversely, in younger, dynamically evolving clusters, the orientations of ETGs tend to reflect their initial infall along filamentary structures, preserving a memory of their accretion path \citep{Plionis02, Plionis03, Rong15b, Wesson84, West94, 2015ApJ...813....6K}. This radial alignment is a known source of bias in weak lensing measurements, where it can mimic the coherent alignments expected from gravitational lensing \citep{Hirata04, Troxel14}.

In large-scale filaments, red ETGs' major axes often align with filament elongations, likely a result of galaxy mergers within these structures \citep{Rong16, Tempel13a, Tempel13b, Tempel15a, Barsanti22, Kraljic21,Kitzbichler03, 2018ApJ...866..138W}. Observations show that redder and brighter ETGs tend to exhibit stronger alignment \citep{Zhang13, Tempel13a}, suggesting that alignment studies can robustly constrain galaxy formation scenarios.

Notably, previous alignment studies primarily focused on red, quiescent ETGs. However, the blue and green ETGs, with younger stellar populations and active star formation, challenge conventional classifications and may offer insights into alternative formation pathways. Their alignment with large-scale structures remains unclear. These ETGs, classfied as non-red ETGs (nETGs), in particular, may exhibit distinct formation mechanisms compared to quiescent ETGs \citep{Tojeiro13, Deng09, Li24}. Examining the alignment of nETGs within large-scale filaments could clarify the differences between red and non-red ETGs, offering a deeper understanding of galaxy evolution and cosmic structure formation.

This study investigates the alignment of nETGs in large-scale filaments. Section~\ref{sec:2} describes the sample selection, Section~\ref{sec:3} presents a statistical analysis of nETG alignments relative to host filaments and compares the findings to red ETG alignments, and Section~\ref{sec:4} provides a summary and discussion of the results.

\section{Sample}
\label{sec:2}

\begin{figure}
   \centering
   \includegraphics[width=\columnwidth, angle=0]{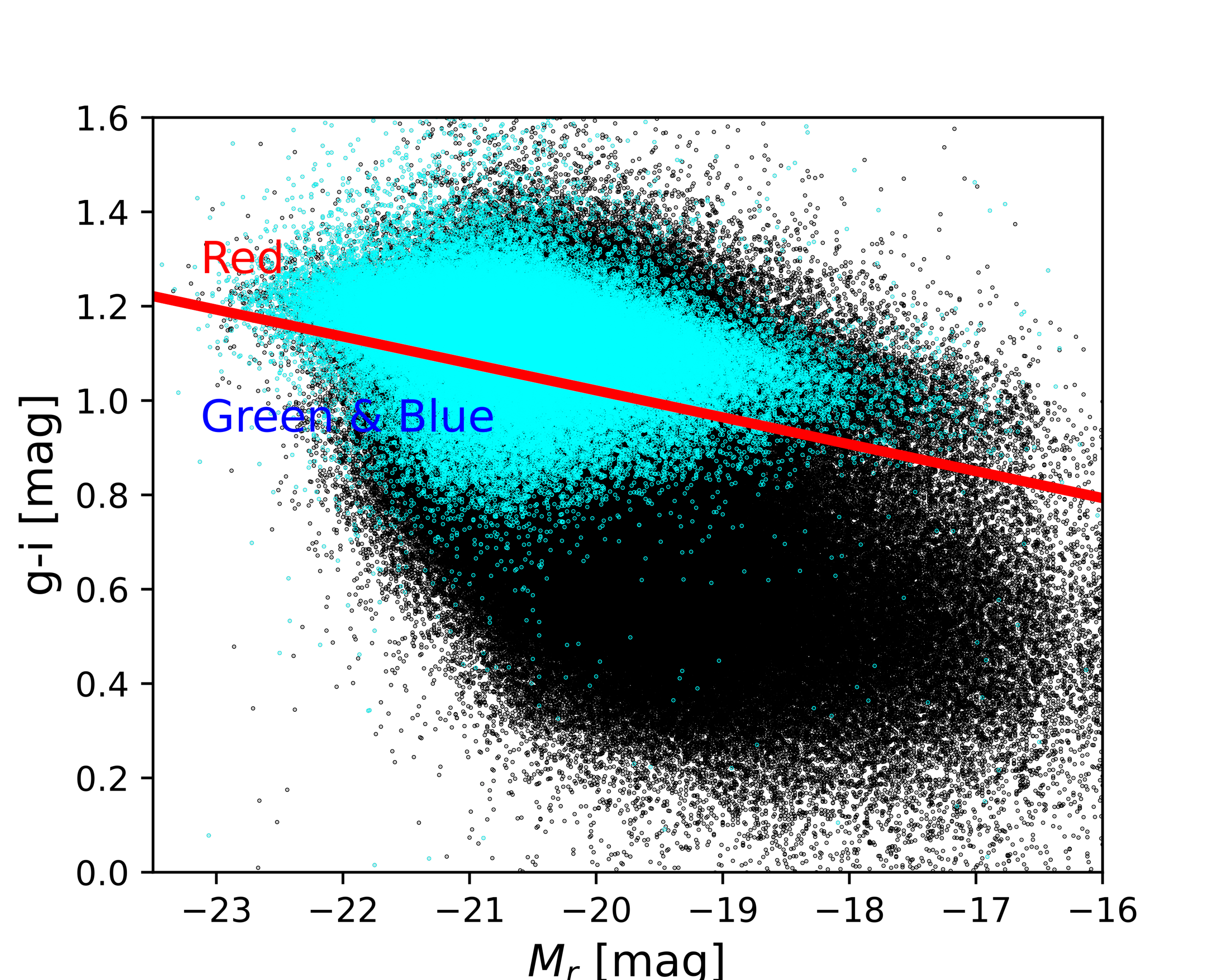}
   \caption{The color-magnitude diagram, showing $g-i$ optical colors versus $r$-band absolute magnitudes $M_r$, for SDSS galaxies (black dots) and ETGs (cyan dots). The red boundary line separating rETGs and nETGs is defined by $g-i=-0.0571(M_r+24)+1.25$ \citep{Papastergis13}.}
   \label{cmd}
   \end{figure}

We select nETGs from the spectroscopic galaxy sample compiled by \cite{Tempel12}, based on data from the Sloan Digital Sky Survey (SDSS) \citep{Aihara11}. This sample is limited to a $r$-band Petrosian magnitude of $m_r = 17.77$~mag, providing various properties for each galaxy, including position, $ugriz$-band absolute magnitudes, morphology (`$morf$'), group ranking (`$rank$'), and more. From the color-magnitude diagram in Fig.~\ref{cmd}, we identify non-red galaxies below the boundary $g-i=-0.0571(M_r+24)+1.25$ \citep{Papastergis13}. Red-sequence galaxies are also selected for comparison.

Galaxies classified as `$morf=2$’ (as defined in \citealt{Tempel12}) are categorized as ETGs, enabling the isolation of both nETGs and red ETGs (rETGs). Stellar masses $M_{\star}$ are estimated from the $r$-band absolute magnitudes and $g-r$ colors using the mass-to-light ratio equation $\log(M_{\star}/L_r) = 1.097(g-r)-0.306$ \citep{Bell03}. Considering the stronger alignment signal in more massive ETGs \citep{Tempel13a, Tempel15a}, only ETGs with $M_{\star}>10^{9.5}\ M_{\odot}$ are included.

Focusing on the alignment of central ETGs within large-scale filaments, we exclude satellite galaxies and retain those with `$rank=1$'. Position angles are determined using the SDSS ``photoObj.deVPhi\_r'' parameter, which is a reliable metric for elliptical galaxies with de Vaucouleurs $R^{1/4}$ surface brightness profile \citep{deVaucouleurs48}. Galaxies with an axis ratio of ``photoObj.deVAB\_r$>0.8$'' are removed for accuracy.

The associated filament of each ETG is identified from the filament catalog by \cite{Tempel14a} based on its three-dimensional distance ($d_{\rm{gf}}$) to the filament spine. The alignment analysis is limited to ETGs within $d_{\rm{gf}}\leq 1.0$~Mpc$/h$, approximately marking a filament boundary \citep{Wang24}. This final sample consists of 968 nETGs and 4,976 rETGs located in their respective host filaments. Their distributions of $M_{\star}$ and $d_{\rm{gf}}$ are compared in the upper panels of Fig.~\ref{fig2}.

%%%%%%%%%%%%%%%%%%%%%%%%%%%%%%%%%%%%%%

\section{Alignment of non-red ETGs}
\label{sec:3}

The catalog compiled by \cite{Tempel14a} includes information on filament points and filament spine orientation at each point. In our study, we calculate the angle $\theta$ between the major axis of each ETG and the orientation of the nearest filament spine in the plane of the sky, following the methodology described by \cite{Wang20}. Here, $\theta$ is constrained to the range [0, 90$^\circ$]. An alignment signal is considered statistically significant if the distribution of $\theta$ deviates noticeably from uniformity.

\begin{figure*}
      \centering
      \includegraphics[width=0.8\textwidth, angle=0]{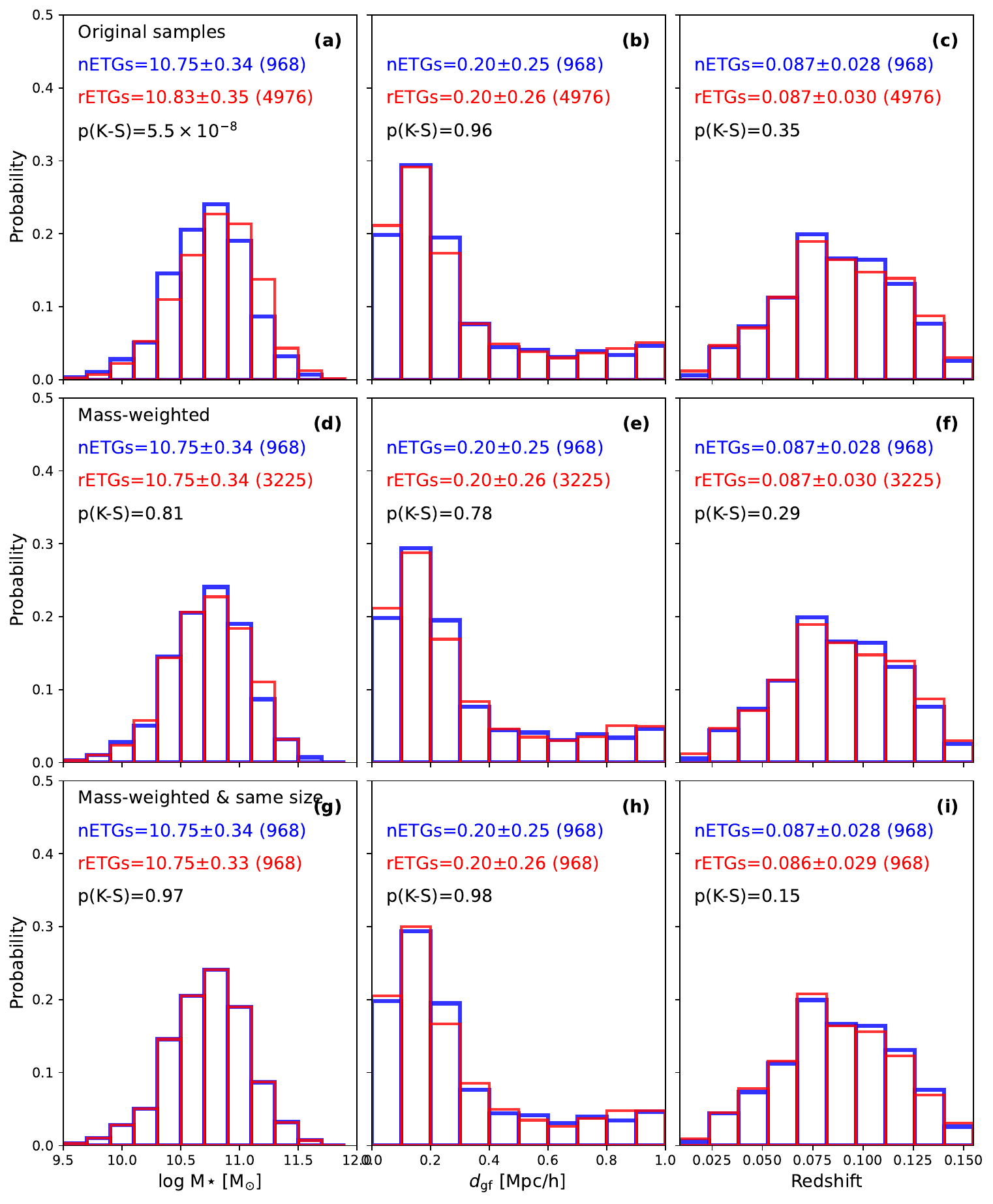}
      \caption{Comparison of the stellar mass distributions (left), $d_{\rm{gf}}$ distributions (middle), and (CMB-corrected) redshift distributions (right) between the nETGs (blue histograms) and rETGs (red histograms). The upper, middle, and lower panels correspond to the results obtained from the original rETG sample, mass-weighted rETG sample, and mass-weighted sample matched in size to the nETG sample, respectively. Each panel displays the median value and standard deviation of the distribution, along with the respective sample sizes in brackets. Additionally, $p$-values from two-sample K-S tests comparing nETGs and rETGs are presented.}
      \label{fig2}
      \end{figure*}

Previous research has shown that the galactic stellar mass, the distance to the filament, and redshift have a significant impact on the strength of the alignment signal of ETGs \citep{Wang20,Chen19,Li13,Donoso06,Faltenbacher09}. As shown in the upper panels of Fig.~\ref{fig2}, while the distributions of $d_{\rm{gf}}$ and redshifts \citep[corrected to the CMB rest
frame;][]{Tempel12} for the rETG and nETG samples are similar, their stellar mass distributions differ. Therefore, it is essential to ensure that the stellar mass distributions of the two samples are comparable to accurately compare their alignment signals. To achieve this, we implement a mass-weighting control method.

      \begin{figure}
      \centering
      \includegraphics[width=0.8\columnwidth, angle=0]{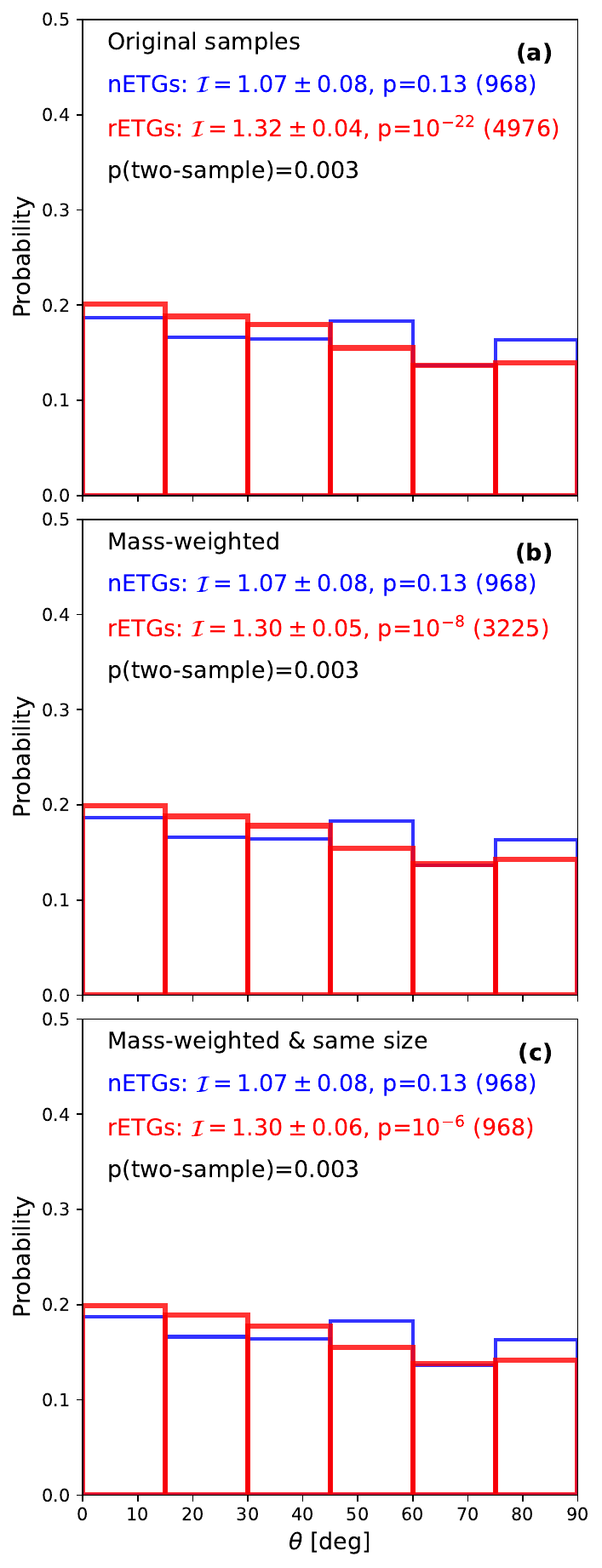}
      \caption{Comparison of the $\theta$ distributions of rETGs (red histograms) and nETGs (blue histograms). The upper, middle, and lower panels display the results from the original rETG sample, mass-weighted sample, and mass-weighted sample with the same size as the nETG sample, respectively. The $1\sigma$ uncertainties, derived from bootstrap resampling while maintaining the sample size (the standard deviation of the distribution functions from 100 repetitions is taken as the uncertainty), are denoted by the colored regions. The uniformly distributed reference is indicated by the gray dashed line. For each sample, the ${\mathcal{I}}(\theta)$ value with its associated uncertainty, K-S test $p$-values compared to a uniform distribution, and sample size (in brackets) are provided. The two-sample K-S test $p$-values for comparing the $\theta$ distributions of rETGs and nETGs are also presented in black in the respective panels.}
      \label{alignment}
      \end{figure}

Initially, we establish the stellar mass distribution of the nETG sample as the reference. Subsequently, we assign weights to the rETG sample using the equation:
\begin{equation}
    w_{{\rm{rETG}},i}=f_{{\rm{nETG}},i}/f_{{\rm{rETG}},i},
\end{equation}
where $f_{{\rm{nETG}},i}$ and  $f_{{\rm{rETG}},i}$ represent the fractions of galaxies in the nETG and rETG samples within the $i$-th stellar mass bin, respectively. $w_{{\rm{rETG}},i}$ denotes the weight of the rETG sample in the $i$-th stellar mass bin. We then randomly select $n'_{{\rm{rETG}},i}$ galaxies from the rETG sample within the $i$-th stellar mass bin, using
\begin{equation}
    n'_{{\rm{rETG}},i}=w_{{\rm{rETG}},i}/{\rm max}(w_{{\rm{rETG}},i})*n_{{\rm{rETG}},i},
    \label{eq2}
\end{equation}
where $n_{{\rm{rETG}},i}$ represents the original number of galaxies in the rETG sample within the $i$-th stellar mass bin. ${\rm max}(w_{{\rm{rETG}},i})$ is the maximum value of $w_{{\rm{rETG}},i}$ across different bins. The stellar mass distribution of the rETG sample post weighting is compared with that of the nETG sample in panel~d of Fig.~\ref{fig2}. The comparable medians and high $p$-values from two-sample Kolmogorov–Smirnov (K-S) tests indicate no significant differences of the stellar masses, $d_{\rm{gf}}$, and redshifts between the two samples after weighting method applied. Subsequently, we utilize the weighted rETG sample (with 3,225 member galaxies) and nETG sample to examine the relationship between their major axes' orientations and filaments' orientations.

The red histogram in panel~b of Fig.~\ref{alignment} depicts the average distribution of $\theta$ of rETGs after mass-weighting, with 100 times repeats (the error bars show the standard deviations of the 100 distribution functions). The major axes of rETGs show clear alignment with the parent filaments. To quantify alignment strength, following the method outlined in \cite{Rong24} and \cite{Rong19}, we define ${\mathcal{I}}(\theta)=N_{0-45}/N_{45-90}$, where $N_{0-45}$ and $N_{45-90}$ represent the numbers of ETGs with $\theta$ in the ranges [0, 45$^{\circ}$] and [45$^{\circ}$, 90$^{\circ}$], respectively. A uniform distribution yields ${\mathcal{I}}(\theta) \simeq 1$.

The uncertainty of ${\mathcal{I}}(\theta)$, $\sigma_{\mathcal{I}}$, is estimated from the 100 times repeats: for each mass-weighting selection of rETGs, we estimate a ${\mathcal{I}}(\theta)=N_{0-45}/N_{45-90}$. The standard deviation of these values is taken as the uncertainty.

The average distribution of $\theta$ for rETGs after mass-weighting reveals a significant alignment signal with ${\mathcal{I}}(\theta)=1.30\pm0.05$ (equivalent to a confidence level of approximately $6\sigma$ based on $({\mathcal{I}}(\theta)-1)/\sigma_{\mathcal{I}}$) and a K-S test $p$-value of $p\sim10^{-8}$ (compared to a uniform distribution). This alignment finding for rETGs is consistent with previous studies \citep[e.g.,][]{Tempel13a}. In contrast, nETGs exhibit no or weak alignment (${\mathcal{I}}(\theta)=1.07\pm0.08${\footnote{The uncertainty of nETGs is estimated using bootstrap resampling. From the original sample with $N$ members, we randomly select $N$ galaxies with replacement, repeating this 100 times to obtain 100 values of ${\mathcal{I}}(\theta)$, and the standard deviation of these values is taken as the uncertainty.}}), with a K-S test $p$-value of $p=0.13$ compared to a uniform distribution, indicating no significant alignment. The K-S test comparison between the rETGs and nETGs yields a small $p$-value ($p\sim 0.003$), confirming a statistically significant difference in their alignment distributions. Therefore, the distinct alignment signals of rETGs and nETGs cannot be attributed to the differences in stellar mass or $d_{\rm{gf}}$ between rETGs and nETGs, as these properties are comparable.

To investigate whether the alignment strength difference between the two samples is influenced by their different sample sizes, we apply the mass-weighting method described earlier and randomly select rETGs by setting $n'_{{\rm{rETG}},i}=n_{{\rm{nETG}},i}$, where $n_{{\rm{nETG}},i}$ represents the number of nETGs in the $i$-th stellar mass bin. The distributions of $M_{\star}$, $d_{\rm{gf}}$, and redshifts of nETGs and the selected rETGs are compared in the lower panels of Fig.~\ref{fig2}, with large $p$-values indicating no significant differences. However, the alignment distributions of the rETGs and selected nETGs with equal sample sizes still exhibit a notable difference. Only the major axes of the rETGs show a significant alignment with filament spines (${\mathcal{I}}(\theta)=1.30\pm0.06$ indicating an alignment signal at approximately $5\sigma$ confidence level; K-S test $p=10^{-6}$ when compared to a uniform distribution). Although the relatively small sample size of nETGs prevents us from entirely ruling out the possibility of a weak alignment between nETGs and their parent filaments, it is evident that the alignment signal strength in nETGs is markedly weaker than that in rETGs. Furthermore, the disparity in alignment signals between nETGs and rETGs is not attributable to differences in sample sizes.

%%%%%%%%%%%%%%%%%%%%%%%%%
\section{Summary and discussion}
\label{sec:4}

Using SDSS data, we investigate the alignment of central ETGs with large-scale filaments. We categorize ETGs into red (rETGs) and non-red (nETGs) based on their optical colors. Our statistical analysis confirms a robust alignment of the major axes of red ETGs with filament orientation, consistent with previous studies. In contrast, nETGs exhibit either no alignment or very weak alignment. This suggests that the alignment signal observed in ETGs within large-scale filaments, as reported in prior research \citep[e.g.,][]{Tempel13a}, is predominantly driven by red ETGs, while blue ETGs do not contribute significantly to the alignment signal.

We establish that the difference in alignment between rETGs and nETGs is not influenced by disparities in the distributions of stellar masses or distances to filament spines between the two samples, nor by variations in sample sizes. Our results remain consistent regardless of changes in ellipticity or stellar mass thresholds for sample selection. For instance, when selecting ETGs based on criteria such as ``photoObj.deVAB\_r$>$0.7'' or ``photoObj.deVAB\_r$>$0.6'' or $M_{\star}>10^{10}\ M_{\odot}$ or $M_{\star}>10^{10.5}\ M_{\odot}$, the alignment patterns remain unchanged.

The absence of alignment in nETGs may indicate a distinct formation pathway. While the alignment of rETGs likely arises from mergers along filament orientations \citep{Tempel13b}, our findings plausibly suggest that nETGs may not predominantly form through mergers. Alternatively, nETGs may arise from processes such as the instability of spiral galaxies \citep{Kormendy13,vandenBosch98}, leading to more randomly distributed orientations. Notably, the star formation histories of nETGs and blue spiral galaxies exhibit similarities \citep{Tojeiro13}.

Furthermore, recent galaxy mergers may have substantially modified the original morphology of nETGs. Studies by \cite{Maller02} indicate that the spin vectors and orientations of galaxies are primarily influenced by their most recent merging events. Therefore, an alternative hypothesis posits that the latest mergers involving nETGs may not align with the filament spine but instead follow more random directions relative to the spine, potentially even perpendicular to it. For example, gas-rich galaxy pairs accreted by large-scale filaments may merge in a manner that positions them on either side of the filament spine, leading to a connecting line between the merging galaxies that is perpendicular to the spine direction. In contrast, mergers giving rise to rETGs are more likely to occur along the spine direction, potentially after undergoing pre-processing within a galaxy group and possessing lower gas content, resulting in a red, quiescent descendant. Therefore, the absence of alignment in nETGs may be attributed to their formation through recently accreted gas-rich galaxy pairs by large-scale filaments, with the merging systems exhibiting more random angular momentum directions.

Alternatively, the most recent mergers of nETGs, which are predominantly found in lower-density environments \citep{Bamford09}, may have occurred more distantly in time compared to rETGs. This scenario may also elucidate the absence or weakness of alignment signals in nETGs, as subsequent gas accretion and star formation may have significantly altered their primordial morphology, resulting in diminished alignment with their parent filaments.

Indeed, establishing the absence of alignment between nETGs and their parent filaments, coupled with probing the underlying causes through cosmological hydrodynamic simulations, constitutes a vital research endeavor. This initiative holds profound implications for advancing our comprehension of the formation and evolution of nETGs. We intend to ameliorate statistical constraints by leveraging imminent large-scale surveys such as the Wide Area VISTA Extragalactic Survey \citep[WAVES;][]{Driver19} and the 4MOST Hemisphere Survey \citep[4HS;][]{Taylor23}, thereby mitigating uncertainties intrinsic to small sample sizes. This approach will enable a more nuanced investigation into the potential alignment, or lack thereof, between nETGs and large-scale filaments from an observational standpoint. Should an absence of alignment be corroborated in expanded nETG cohorts, it would fundamentally challenge prevailing models of nETG genesis and development. We envisage employing sophisticated cosmological hydrodynamic simulations, such as Illustris-TNG \citep{Nelson19}, to delve deeper into the intricate formation processes of nETGs.

%%%%%%%%%%%%%%%%%%%%%%%%%%%%%%%%%%%%%%%%%%%%%%%%%%%%%%%%%%%%%%%%%%%%%%%%%%%%%%%%
%\clearpage
\acknowledgments

Y.R. acknowledges supports from the CAS Pioneer Hundred Talents Program (Category B), the NSFC grant 12273037, the USTC Research Funds of the Double First-Class Initiative, and the research grants from the China Manned Space Project (the second-stage CSST science projects: ``Special Galaxies Studies based on CSST NUV and Multi-wavelengh Surveys'' and ``A Study of Small-scale Structures of Galaxies \& observational predictions from simulations''). P.W. sponsored by Shanghai Pujiang Program(No.22PJ1415100), Shanghai Rising-Star Program (No.24QA2711100). P.W. acknowledge financial support by the NSFC (No. 12473009).

%\end{acknowledgments}
%%%%%%%%%%%%%%%%%%%%%%%%%%%%%%%%%%%%%

%\clearpage
%%%%%%%%%%%%%%%%%%%%%%%%%%%%%%%%%%%%%
%\bibliographystyle{aastex62}

%%%%%%%%%%%%%%%%%%%%%%%%%%%%%%%%%%%%%%%%%%%%%%%%%%

\end{document}